\documentstyle[prb,floats,aps,amssymb,eqsecnum,epsf,rotate]{revtex}
\begin{document}

\draft

\title{Spin waves and phase diagram of one-dimensional, dipolar 
antiferromagnets}
\author{M. Hummel, C. Pich,\thanks{Present address: Physics Department,
University of California, 
Santa Cruz, CA 95064.} and F. Schwabl}
\address{Institut f\"ur Theoretische Physik, Physik-Department der Technischen
Universit\"at M\"unchen, James-Franck-Strasse, 85747 Garching, Germany}

\date{\today}

\maketitle

\begin{abstract}
We report on the properties of a dipolar, antiferromagnetic chain in the 
framework of linear spin-wave theory. As a model we use an isotropic
Heisenberg-Hamiltonian with antiferromagnetic nearest neighbor exchange 
and anisotropic dipole-dipole interaction. The phase diagram is calculated
for an arbitrary ratio of dipolar interaction to exchange interaction
and for fields both parallel and perpendicular to the chain direction.
We can distinguish two regions, an antiferromagnetic exchange dominated region,
in which the dipole energy acts as a planar anisotropy and a ferromagnetic
dipole dominated region, in which the dipole energy acts as an easy-axis
anisotropy. The anisotropy in the field dependent magnetization, which is 
caused by quantum fluctuations, is several orders of magnitude larger than the 
anisotropy predicted by the classical theory. We show that the dipole-dipole 
interaction does not lead to long-range order in the pure one-dimensional 
case, however for quasi one-dimensional systems spin-waves are important. 
The calculated magnetization is 
compared with experimental measurements for $\rm{RbMnBr_3}$, a quasi 
one-dimensional antiferromagnet.
\end{abstract}

%insert suggested PACS numbers in braces on next line
\pagebreak
\narrowtext
\twocolumn

%\begin{multicols}{2}
\narrowtext

\section{Introduction}

Low-dimensional magnetic systems have long attracted the attention of
experimental and theoretical physics. In systems with reduced 
dimensionality the fluctuations are enhanced and can prevent a system from
long-range ordering. For one- and two-dimensional isotropic
Heisenberg ferro- and antiferromagnets with short-range interaction it has been
proven exactly \cite{Hohen67,Mermin66} that long-range order is absent for
finite temperatures. The existence of long-range order in low-dimensional
systems depends crucially on anisotropies and long-range interactions, as there
are single-ion anisotropy or the dipole-dipole interaction. 

In real systems the dipole-dipole interaction is always present in addition to
the short-range exchange interaction. Although the dipole-dipole energy is weak
compared to the exchange energy it plays an important role in low-dimensional
systems due to its anisotropic and long-range character. Maleev \cite{Male76}
showed that in a two-dimensional ferromagnetic system the additional dipolar
interaction leads to a modification of the spin-wave frequency in the long
wavelength limit from a quadratic to a square root wave vector dependence, for
which long-range order is established. For the antiferromagnetic counterpart on
a square and a honeycomb lattice it has also been shown \cite{pich93} that
dipolar interaction stabilizes long-range magnetic order at finite
temperatures. In this cases, however, the dipolar anisotropy leads to a
discrete ground state with spins oriented perpendicular to the plane,
producing an energy gap in the dispersion relation. The finite N\'eel
temperatures in various quasi two-dimensional systems (e.g. K$_2$MnF$_4$) have
been calculated within the dipolar model \cite{pich93}, in good agreement with
the experimental values.

In recent papers \cite{pich95a} we determined the magnetic phase
diagram of the two-dimensional, dipolar Heisenberg antiferromagnet (square and
honeycomb lattice) for fields along the easy-axis, i.e. perpendicular to the
plane. In contrast to models with single-ion anisotropy or an anisotropic
exchange interaction, we found that the dipole-dipole interaction establishes a
new phase, an intermediate phase. With increasing field strength the system
changes by second-order transitions from the N\'eel phase to an intermediate
phase, a spin-flop phase and finally to a paramagnetic phase at high magnetic
fields.

From the results in two dimensions we anticipate that also in quasi
one-dimensional Heisenberg antiferromagnets, the dipolar interaction
will reduce thermal and quantum fluctuations and thus increase
the tendency to order.

In this paper we investigate the influence of the dipolar interaction on the 
classical ground states and on quantum fluctuations. It is shown that due to 
the competition of the exchange and dipolar interaction we can distinguish two
regions, namely an exchange dominated and a dipolar dominated region.
In the former the dipolar energy leads to a planar anisotropy, whereas
in the latter it leads to an easy-axis anisotropy. We study
the dispersion relations for various relative strengths of the dipolar 
interaction. In the exchange dominated region, the dipolar
interaction breaks the continuous symmetry of the ground state partially 
and favors states with spins perpendicular to the chain direction. Rotation
symmetry around the chain direction still exists. The dispersion
relation for this phase is modified by the dipolar interaction, 
but as in the isotropic counterpart no long-range order can exist even for 
$T=0$.

The most famous quasi one-dimensional systems are
the ternary compounds ABX$_3$ (A alkaline, B transition metal and X halogen),
which have been studied intensively experimentally in the context of 
Haldane's phase \cite{Haldane} and solitonic excitations \cite{Steiner91}. 
Of particular interest are the Mn compounds. Because the angular momentum 
$L$ is zero, no crystal field splitting occurs in these systems and the 
dipole-dipole interaction should be the most important anisotropy.
Experiments \cite{abanov} show an anisotropy in the magnetization
measurements for fields applied parallel and perpendicular to the spin-chain 
axis. We will show that this anisotropy can be understood qualitatively by our 
simple one-dimensional dipolar model.
Up to now, only partial aspects of the dipole-dipole 
interaction have been considered \cite{Shiba,Santini}. 
In Ref[9] the classical ground state has been determined, but no
excitations have been
investigated, whereas in Ref[10] the dipolar interaction is only 
considered for nearest neighbors. In this paper the effects of the additional
dipolar interaction will be studied in a systematic way.

In this spin wave theory we disregard kink-like excitations, which even for the
$r^{-3}$ decay of the dipolar interaction destroy one-dimensional long-range
order. However, for wavelengths shorter than the
correlation length magnon excitations become important. This has been shown 
experimentally by Steiner for CsNiF$_3$ \cite{Steiner}. The short range
antiferromagnetic order supports spin waves with wave vectors not too close to
the center of the Brillouin zone. Especially for quasi one-dimensional 
systems the weak interchain interaction leads to a stabilization of the 
magnetic order and spin waves become the relevant excitations.

The outline of this paper is as follows: in section II we introduce the dipolar
model; in III we investigate the ground state, the spin-wave spectrum and the
order parameter for vanishing fields; in IV we study the magnetic phase
diagram for the whole parameter region. In the last section we apply our model
to a quasi one-dimensional system, namely RbMnBr$_3$.

\section{Model}
The Hamiltonian of a dipolar antiferromagnet reads
\begin{equation}
H=-\sum_{l \neq l'} \sum_{\alpha \beta} (J_{ll'} \delta_{\alpha \beta}
+A_{ll'}^{\alpha\beta}) S_l^{\alpha} S_{l'}^{\beta} 
- g \mu_B {\bf{H}}_0 \sum_l {\bf{S}}_l \;,
\label{Hamiltonian}
\end{equation}
with spins ${\bf{S}}_l$ at lattice sites ${\bf{x}}_l$. 
The first term in brackets is the isotropic exchange interaction $J_{ll'}$ 
and the second term is the classical dipole-dipole-interaction
\begin{eqnarray}
A_{ll'}^{\alpha\beta}= -\frac{1}{2} {(g \mu_B)^2} &\Biggl(&
\frac{\delta_{\alpha\beta}}{{|{\bf{x}}_l-{\bf{x}}_{l'}|}^3}\nonumber\\
&-& \frac{3 ({{\bf{x}}_l-{\bf{x}}_{l'})}_{\alpha} 
{({\bf{x}}_l-{\bf{x}}_{l'})}_{\beta}}{{|{\bf{x}}_l-{\bf{x}}_{l'}|}^5}
\Biggr) \, .
\label{ddw}
\end{eqnarray}
The third term in Eq. (\ref{Hamiltonian}) describes the effect of an 
external magnetic field ${\bf{H}}_0$,
where $g$ is the Land\'{e}-factor and $\mu_B$ is the Bohr magneton.
 From this Hamiltonian we calculate the magnon spectrum in the framework of the
Holstein-Primakoff (HP)\cite{HP} transformation. 
The Holstein-Primakoff transformation is an
exact transformation which expresses the Hamiltonian of Eq. (\ref{Hamiltonian})
in terms of Bose operators $a_l^{\dagger}, a_l$. For low
temperatures an expansion up to bilinear terms can be used
\begin{eqnarray}
\tilde{S}_{l}^{x} &= &
\sqrt{\frac{S}{2}} (a_l + {a}_{l}^{\dagger}) \; , \quad
\tilde{S}_{l}^{y} = -i \sqrt{\frac{S}{2}} (a_l - {a}_{l}^{\dagger}) \; ,
\nonumber\\
 & & \label{HP}\\
\tilde{S}_{l}^{z} &=& S - {a}_l^{\dagger} a_{l} \quad .\nonumber 
\end{eqnarray}
The tilde indicates that we have to use these inner spin coordinates in the
rotated coordinate frame where the $z$-direction lies in the direction of the 
classical spins. 
For the following discussion we define a dimensionless parameter $\kappa$, 
which gives the ratio between the dipolar energy and the exchange energy:
\begin{equation}
\kappa=\frac{{(g \mu_B)}^2}{c^3 J} \ .
\label{kappa}
\end{equation}
Here $J$ denotes the strength of the exchange energy (the negative sign of
the antiferromagnetic exchange interaction $J_{ll'}$ is taken into account 
separately) and $c$ denotes the distance between magnetic ions.\\
In the following we consider a spin chain which is directed along the $x$-axis
(see Fig. \ref{fig1}a,b).

\section{Dipolar antiferromagnetic chain without field}

The long-range dipole-dipole interaction and the antiferromagnetic exchange
interaction compete with each other: while the latter favors an
antiferromagnetic spin configuration, the former favors a
ferromagnetic orientation along the chain axis. Thus, we can distinguish an
exchange-dominated region and a dipole-dominated region,
which are separated by the condition (to be shown in Eq. (\ref{bed1}))
\begin{equation}
J_{{\bf{q}}_0}-J_0 +A_{{\bf{q}}_0}^{33} - A_0^{11}\; \lessgtr \; 0
 \; ,
\label{condition}
\end{equation}
where $J_{\bf{q}}$ is the Fourier transform of the 
exchange interaction $J_{ll'}$, which equals $J_{\bf{q}}=-2J \cos (q_x c)$ 
for nearest-neighbor interaction (in the following we set $q_x=q$), 
and $A_{\bf{q}}^{\alpha\beta}$ 
is the Fourier transform of the dipolar tensor Eq. (\ref{ddw}) 
\begin{equation}
A_{\bf{q}}^{\alpha \beta} = \sum_{l \neq 0} A_{l 0}^{\alpha \beta} 
\exp{(i {\bf{q}} {\bf{x}}_l)} \;,
\label{dipft}
\end{equation}
which can be evaluated by Ewald summation \cite{Ewald} (see appendix A). The
wave vector ${\bf{q}}_0=\frac{\pi}{c}(1,0,0)$ describes the antiferromagnetic
order. Eq. (\ref{condition}) leads to a critical value of $\kappa$
\begin{equation}
\kappa_c=\frac{4}{2\zeta(3)-\eta(3)} \approx 2.66 \;,
\label{criticalkappa}
\end{equation}
with $\zeta(n)$ and $\eta(n)$ defined in appendix A.

In the following we investigate the classical ground states and
excitation spectra of the one-dimensional chain as a function of $\kappa$.

\subsection{Exchange dominated region ($\bbox{\kappa < \kappa_c}$)}
 
\begin{figure}[htb]
  \narrowtext 
 \epsfxsize=0.6\columnwidth\rotate[r]{
  \epsfbox{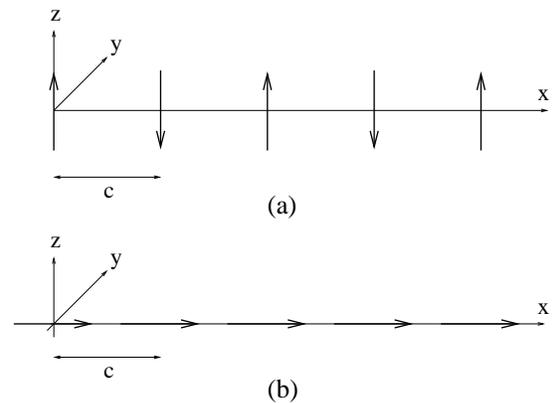}}
\hspace{1.5cm}
  \caption{a) Classical ground state of the dipolar, antiferromagnetic chain 
in the exchange dominated region ($\kappa < \kappa_c$). 
b) Ground state for the dipolar dominated region ($\kappa > \kappa_c$).}
\label{fig1}
\end{figure}

For pure exchange energy $J$, the classical ground state is the
configuration where the spins are antiparallel, as described by the
antiferromagnetic wave vector ${\bf{q}}_0$.
This state is continuously degenerate due to the rotational symmetry.
The dipolar interaction breaks the rotational symmetry and favors the
antiferromagnetic ground state, in which the spins lie in a plane perpendicular
to the chain axis, as can be seen by the largest value of the dipole-dipole
tensor ($A_{{\bf{q}}_0}^{33} = A_{{\bf{q}}_0}^{22} > A_{{\bf{q}}_0}^{11}$). 
However,
rotation symmetry around the $x$-axis still exists (s.Fig. \ref{fig1}a).

\begin{figure}[htb]
  \narrowtext 
 \epsfxsize=0.6\columnwidth\rotate[r]{
  \epsfbox{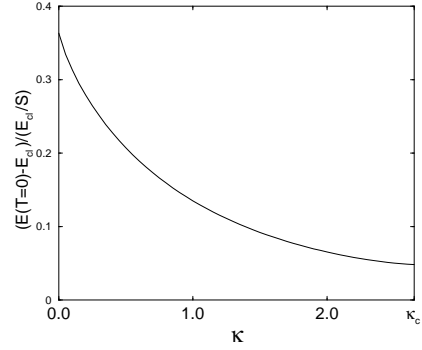}}
\hspace{1cm}
  \caption{Deviation of the ground state energy from the classical 
value due to quantum fluctuations for $\kappa < \kappa_c$. For 
$\kappa \geq \kappa_c$ there is no difference between the two energies, 
because the Hamiltonian is already diagonal after the HP-transformation 
(s. Eq. ({\protect{\ref{2.22}}})).}
\label{energy}
\end{figure}

In the following calculations the primitive cell is the chemical cell which is
half the magnetic unit cell. 
The antiferromagnetic modulation is taken into account via
\begin{equation}
\sigma_l=e^{i {\bf{q}_0} {\bf{x}}_l} = \left\{
\begin{array}{ll}
+1& l \in {\cal{L}}_1 \\
-1& l \in {\cal{L}}_2
\end{array}
\quad , \quad {\bf{q}}_0=\frac{\pi}{c}(1,0,0)
\label{modulation}
\right.
\end{equation}
where ${\cal{L}}_1 ({\cal{L}}_2)$ denotes the first (second)
sublattice\cite{Ziman}.
If we start from the classical ground state with the spins oriented along the
$\pm z$-axis and perform the Holstein-Primakoff transformation, we get the 
following Fourier transformed Hamiltonian
\begin{equation}
H=E_{cl} + \sum_{\bf{q}} A_{\bf{q}}
a_{\bf{q}}^{\dagger} a_{\bf{q}} + \frac{1}{2} B_{\bf{q}}
(a_{\bf{q}} a_{-\bf{q}} + a_{\bf{q}}^{\dagger} a_{-\bf{q}}^{\dagger})
\label{hamvordiag}
\end{equation}
with the classical ground state energy
\begin{equation}
E_{cl} = - NS^2 (J_{{\bf{q}}_0} + A_{{\bf{q}}_0}^{33})
\end{equation}
and the coefficients
\begin{eqnarray}
A_{\bf{q}} & = & S(2J_{{\bf{q}}_0}-J_{{\bf{q}}+{\bf{q}}_0}-J_{\bf{q}})
+S(2 A_{{\bf{q}}_0}^{33} - A_{{\bf{q}}+{\bf{q}}_0}^{22} - A_{\bf{q}}^{11})
\nonumber\\
B_{\bf{q}} & = & S (J_{{\bf{q}}+{\bf{q}}_0}-J_{\bf{q}})
+S (A_{{\bf{q}}+{\bf{q}}_0}^{22}-A_{\bf{q}}^{11}) \;.
\label{bqspectrum}
\end{eqnarray}
The Hamiltonian (\ref{hamvordiag}) is diagonalized by a Bogoliubov 
transformation\cite{Akhiezer} and reads in terms of the new Bose creation and 
annihilation operators $b_{\bf{q}}^{\dagger}$ and $b_{\bf{q}}$
\begin{equation}
H = E_{cl} + \frac{1}{2} \sum_{\bf{q}} (E_{\bf{q}}-A_{\bf{q}}) + \sum_{\bf{q}}
E_{\bf{q}} \, b_{\bf{q}}^{\dagger} b_{\bf{q}} \; .
\label{hamdiag}
\end{equation}
The second term in Eq. (\ref{hamdiag}) is a correction to the ground state 
energy due to quantum fluctuations, while the third term 
is the harmonic magnon Hamiltonian.
The ground state energy
\begin{equation}
E(T=0)=E_{cl} + \frac{1}{2} \sum_{\bf{q}} (E_{\bf{q}}-A_{\bf{q}})
\end{equation}
is shown in Fig. \ref{energy} as a function of $\kappa$. The quantum
fluctuations are strongest for pure exchange interaction \cite{Anderson}. 
The deviation of the ground state energy from the classical value
\begin{equation}
E(T=0)=E_{cl}(1+0.363/S),
\end{equation}
is most significant in this case
and decreases with increasing dipolar interaction, i.e. the dipolar interaction
tends to stabilize the antiferromagnetic ground state.

The magnon dispersion relation reads
\begin{eqnarray}
E_{\bf{q}}&=&\sqrt{A_{\bf{q}}^2-B_{\bf{q}}^2}\nonumber\\
&=&2S\sqrt{(J_{{\bf{q}}_0}
-J_{\bf{q}}+A_{{\bf{q}}_0}^{33}-A_{\bf{q}}^{11})}\nonumber\\
&&\times \sqrt{(J_{{\bf{q}}_0}-J_{{\bf{q}}+{\bf{q}}_0}+A_{{\bf{q}}_0}^{33}
-A_{{\bf{q}}+{\bf{q}}_0}^{22})} \; ,
\label{magspectrum}
\end{eqnarray}
\begin{figure}[htb]
  \narrowtext
 \epsfxsize=0.6\columnwidth\rotate[r]{
  \epsfbox{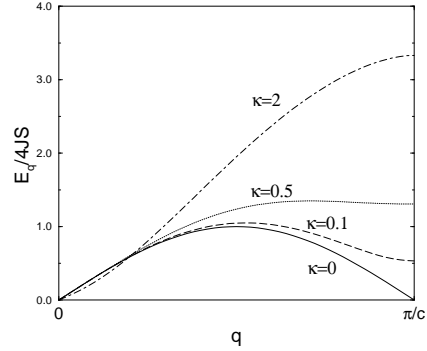}}
\mbox{}
\hspace{4cm}
  \caption{Dispersion relation for different values of $\kappa$
in the exchange dominated region ($\kappa \leq \kappa_c$).}
\label{gskleink}
\end{figure}
which is shown in Fig. \ref{gskleink} for four values of $\kappa$. 
Because of the continuous degeneracy of the ground state 
(rotation around the $x$-axis) the Goldstone-mode at ${\bf{q}}=0$
survives in the presence of the dipolar interaction, 
but the behavior for small wave vectors is modified.
 If we expand the magnon spectrum around the center of the Brillouin zone, 
we get logarithmic corrections 
\begin{eqnarray}
E_{\bf{q}}/(4JS) &=& \biggl\{  b_1 {(qc)}^2 
-\frac{1}{4} \kappa \Bigl( 1+ \frac{1}{2} \kappa c_5 \Bigr) 
{(qc)}^4 \log {(qc)} \nonumber\\
&&{+b_2 {(qc)}^4 +{\cal{O}} \Bigl( {(qc)}^6 \Bigr) \biggr\}}^{1/2}\;,
\label{entw}
\end{eqnarray} 
with $q$-independent constants $c_5,b_1,b_2$ (see Eq. 
\ref{cvalues}, \ref{bvalue1}, \ref{bvalue2} in Appendix A).

The order parameter, the staggered magnetization, is given by
\begin{equation}
{\text{N}} = g \mu_B \sum_l \sigma_l \langle \tilde{S}_l^z \rangle
\end{equation}
with $\sigma_l$ from Eq. (\ref{modulation}). 
Inserting the Holstein-Primakoff transformation (\ref{HP}), we obtain
\begin{equation}
{\text{N}} = g \mu_B N S - {\text{N}}_0 - {\text{N}}_{th}(T) 
\end{equation}
with the zero-point contribution
\begin{equation}
{\text{N}}_0 
= \frac{1}{2} g \mu_B \sum_{\bf{q}} \left( \frac{A_{\bf{q}}}{E_{\bf{q}}} -1
\right) 
\label{2.16}
\end{equation}
and
\begin{equation}
{\text{N}}_{th}(T) = 
g \mu_B \sum_{\bf{q}} \frac{A_{\bf{q}}}{E_{\bf{q}}} \, n_{\bf{q}} \;,
\end{equation}
from thermal excitations of spin waves. 
$n_{\bf{q}}=\frac{1}{e^{E_{\bf{q}}/kT}-1}$ 
denotes the Bose-occupation number. Both quantum and thermal
fluctuations reduce the order parameter.
By use of the small wave vector expansion, Eq. ({\ref{entw}}), we see 
that the sum ${\text{N}}_0$ diverges for ${\bf{q}} \rightarrow 0$.
Even though the dipolar interaction has the tendency to stabilize 
the antiferromagnetic order, it is nevertheless destroyed by quantum 
fluctuations even at $T=0$, as found for the isotropic counterpart.
There is yet another stability condition resulting from the requirement that
the spin wave frequencies (Eq. (\ref{magspectrum})) be real which leads to
\begin{eqnarray}
\label{aqbed}
A_{\bf{q}} &>& 0\\
\label{bqbed}
A_{\bf{q}} &>& |B_{\bf{q}}|
\end{eqnarray}
for all wave vectors $\bf{q}$\cite{Cohen}. 
We get a lower bound for the strength of the exchange energy from the first 
condition (Eq. (\ref{aqbed})) at ${\bf{q}}=0$
\begin{equation}
J_{{\bf{q}}_0}-J_0+A_{{\bf{q}}_0}^{33}-A_0^{11} > 0 \;.
\label{bed1}
\end{equation}
For lower exchange energies the system changes to the dipolar dominated
region.

\subsection{Dipolar dominated region ($\bbox{\kappa > \kappa_c}$)}

For pure dipolar interaction the classical ground state consists of two 
states, in which the spins are oriented ferromagnetically along the $\pm
x$-axis (Fig. \ref{fig1}b), because the largest eigenvalue of the dipolar
tensor is $A_{0}^{11}$ (s. Fig. \ref{fig12}), i.e. the dipolar interaction acts
as an easy-axis anisotropy.
If we start with this ground state and perform the HP-transformation we get
\begin{equation}
H=E_{cl} + \sum_{\bf{q}} E_{\bf{q}} a_{\bf{q}}^{\dagger} a_{\bf{q}} \;,
\label{2.22}
\end{equation}
with the classical ground state energy
\begin{equation}
E_{cl} = - NS^2 (J_0 + A_{0}^{11})
\end{equation}
and
\begin{equation}
E_{\bf{q}} = 2S (J_0 - J_{\bf{q}}) + 2S (A_0^{11} - A_{\bf{q}}^{33}) \ .
\label{ferrospectrum}
\end{equation}
Due to the broken rotation symmetry the spectrum has an energy gap at the
zone center
\begin{displaymath}
E_0 = 2S (A_0^{11} - A_0^{33}) \;.
\end{displaymath}
\begin{figure}[htb]
  \narrowtext
 \epsfxsize=0.6\columnwidth\rotate[r]{
  \epsfbox{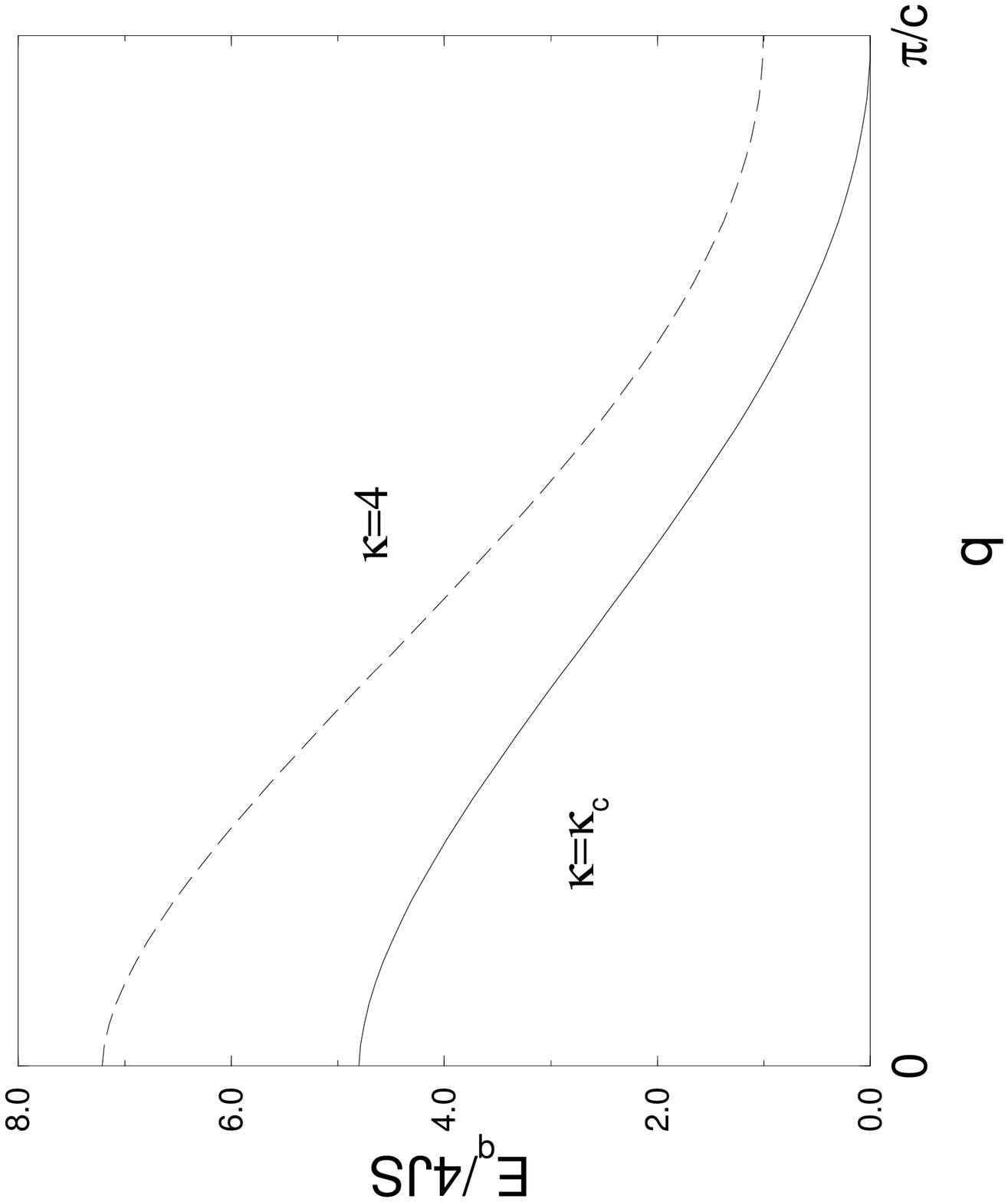}}
\mbox{}
\hspace{4cm}
  \caption{Spin-wave dispersion relation of the dipolar, antiferromagnetic
chain for $\kappa =\kappa_c=2.66$ (solid) and $\kappa =4$ (dashed).}
   \label{dispgap}
\end{figure}
The minimum of the dispersion relation is found at the zone edge (s. Fig. 4),
which vanishes for $\kappa=\kappa_c$ and increases with increasing dipole
strength. 

In linear spin-wave theory the classical ground state and the quantum
mechanical ground state coincide. The Hamiltonian (\ref{2.22}) is diagonalized
by the HP-trans\-formation, and there are no corrections for quantum
fluctuations\cite{pich2000} at $T=0$, i.e. the ground state has long-range
order.  The order parameter is the magnetization in $x$-direction:
\begin{equation}
M^x =  g \mu_B \sum_l \langle S_l^x \rangle =
g \mu_B N S - M_{th}(T)
\label{2.14}
\end{equation}
with
\begin{equation}
M_{th}(T)=g \mu_B \sum_{\bf{q}} \frac{1}{e^{E_{\bf{q}}/k_B T}-1} \ .
\label{mtherm}
\end{equation}
For $\kappa > \kappa_c$ the spin-wave spectrum has gaps at ${\bf{q}}=0$ and 
${\bf{q}}={\bf{q}}_0$ (Eq. (\ref{ferrospectrum})). Thus, the sum in 
Eq. (\ref{mtherm}) is finite for finite
temperatures and from point of view of the linear spin-wave theory long-range
order should be possible. However, as was shown by
Dyson \cite{Dyson}, kink-like thermal excitations destroy the long-range
order in ferromagnetic Ising systems with interactions decreasing like 
$J(r) \propto \frac{1}{r^\alpha}$ for $\alpha > 2$. Thus, we expect that for a
Heisenberg system with dipole-dipole interaction these nonlinear excitations 
will ultimately destroy the long-range order as well. 

Stability of the ground state requires that the spectrum is positive,
$E_{\bf{q}}>0$. For decreasing dipole energy the instability of the spectrum 
sets in at ${\bf{q}}_0$, from which we recover the stability condition
Eq. (\ref{condition})
\begin{equation}
J_{{\bf{q}}_0}-J_0+A_{{\bf{q}}_0}^{33}-A_0^{11} < 0 \;.
\label{bed2}
\end{equation}
The soft mode at ${\bf{q}}_0$ for $\kappa=\kappa_c$
indicates the phase transition to the antiferromagnetic phase
(exchange-dominated region).

Consequently at $T=0$ there is long-range order (LRO) in the dipolar dominated
region, in contrast to the exchange dominated region. However, for finite
temperatures no LRO exists at all due to nonlinear excitations, namely domain
walls.

\section{Magnetic phase diagram}

In this section we determine the magnetic phase diagram of the
dipolar antiferromagnetic chain for both the dipolar dominated and the exchange
dominated region. We study homogeneous magnetic fields along and perpendicular
to the chain axis ($x$-direction). The results are presented in terms of
the parameter $\kappa$ (Eq. (\ref{kappa})) and the dimensionless field 
$h=\frac{g \mu_B H_0}{JS}$. The spin orientations are determined from
the classical ground state energy within a two-sublattice model.

\subsection{Field in chain direction}
First we study the case of a magnetic field along the chain axis. In the
dipolar region ($\kappa > \kappa_c$) the field is parallel to the spins, and
thus no reorientation takes place.

\begin{figure}[htb]
  \narrowtext
 \epsfxsize=0.6\columnwidth\rotate[r]{
  \epsfbox{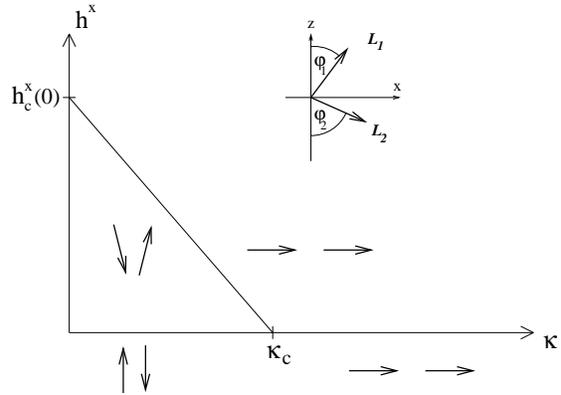}}
\mbox{}
\hspace{4.5cm}
  \caption{Magnetic phase diagram of the dipolar, antiferromagnetic chain with
  a field applied in chain direction. The spin arrangements below the $x$-axis
  indicate the phases for vanishing field. In the insert the
  configuration of the spins in a two-sublattice model is shown.}
\label{phasex}
\end{figure}

\begin{figure}[htb]
  \narrowtext
 \epsfxsize=0.6\columnwidth\rotate[r]{
  \epsfbox{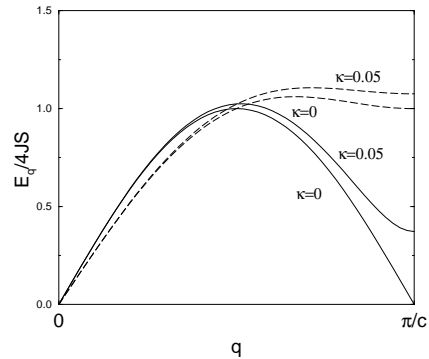}}
\mbox{}
\hspace{5cm}
\caption{Dispersion relation for the linear chain in the spin-flop phase for
  two values of $\kappa$; solid lines for vanishing field and
  dashed  lines for $\sin \varphi=0.5$ (Eq. ({\protect{\ref{phix}}})).}
\label{dispx}
\end{figure}

In the exchange dominated region ($\kappa < \kappa_c$) the
spins are oriented perpendicular to the $x$-axis and any infinitesimal field
along the axis will lead to a reorientation of the spins in order to gain 
energy from the Zeeman term (Eq. (\ref{Hamiltonian})). 
In a two-sublattice model (the spins of the sublattice ${\cal{L}}_1
({\cal{L}}_2)$ enclose an angle $\varphi_1 (\varphi_2)$ with the $y$-$z$-plane 
(see Fig. \ref{phasex})) we can calculate the classical ground state energy. 
For spins lying in the $xz$-plane we get
\begin{eqnarray}
E_{cl}=-\frac{NS^2}{4} &\Bigl[ &
(J_{0}+A_{0}^{11}) {(\sin \varphi_1 + \sin \varphi_2)}^2 \nonumber\\
&+&(J_{{\bf{q}}_0}+A_{{\bf{q}}_0}^{11}) {(\sin \varphi_1 - \sin
\varphi_2)}^2 \nonumber\\
&+&(J_0+A_{0}^{33}) {(\cos \varphi_1 - \cos \varphi_2)}^2 \nonumber\\
&+&(J_{{\bf{q}}_0}+A_{{\bf{q}}_0}^{33}) {(\cos \varphi_1 + \cos \varphi_2)}^2
\Bigr] \nonumber \\
&-&\frac{1}{2}g \mu_B H_0^x N S (\sin \varphi_1+\sin \varphi_2)\;.
\label{eclassicx}
\end{eqnarray}
Minimization with respect to $\varphi_1$ and $\varphi_2$ yields
$\varphi_1=\varphi_2=\varphi$, i.e. a usual spin-flop phase. From the minimum
one determines the relation between the angle $\varphi$ and the magnetic 
field:
\begin{equation}
\sin \varphi = \frac{g \mu_B H_0^x}{2S(J_{{\bf{q}}_0}-J_0+A_{{\bf{q}}_0}^{33}
-A_0^{11})} \ . \label{phix}
\end{equation}
Thus the spins become progressively aligned along the field as the field
increases. There is a continuous transition from this spin-flop phase to the 
paramagnetic phase for $\varphi=90^\circ$. 
The transition occurs at
\begin{equation}
h_{c}^x (\kappa) = 8 + \kappa \bigl( 2 \eta(3) - 4 \zeta(3) \bigr)
\quad {\text{for}} \quad \kappa<\kappa_c \; .
\end{equation}
The complete phase diagram is shown in Fig. \ref{phasex}.
The HP-transformed Hamiltonian in the spin-flop phase has the form of
Eq. (\ref{hamvordiag}) with coefficients 
\begin{eqnarray}
A_{\bf{q}}&=& S(2J_{{\bf{q}}_0}-J_{{\bf{q}}+{\bf{q}}_0}-J_{{\bf{q}}}
+ 2 A_{{\bf{q}}_0}^{33} - A_{{\bf{q}}+{\bf{q}}_0}^{22} 
- A_{\bf{q}}^{11})\nonumber\\
&&+S \sin^2 \varphi \, 
(J_{\bf{q}}-J_{{\bf{q}}+{\bf{q}}_0}+A_{\bf{q}}^{11}-
A_{{\bf{q}}+{\bf{q}}_0}^{33})\nonumber\\
B_{\bf{q}}&=& S(J_{{\bf{q}}+{\bf{q}}_0}-J_{\bf{q}}
+A_{{\bf{q}}+{\bf{q}}_0}^{22} 
- A_{\bf{q}}^{11})\nonumber\\
&&+S \sin^2 \varphi \, (J_{\bf{q}}-J_{{\bf{q}}+{\bf{q}}_0}
+A_{\bf{q}}^{11}-A_{{\bf{q}}+{\bf{q}}_0}^{33}) \;,
\label{aqbqx}
\end{eqnarray}
where $\varphi$ is related to the field by Eq. (\ref{phix}).
The dispersion relation is shown in Fig. \ref{dispx} for two values 
of the field.

The Goldstone-mode at ${\bf{q}}=0$ remains for finite dipolar energy 
because the system is still invariant for rotations around the $x$-axis.

The magnetization in the spin-flop phase can be calculated from the free
energy $F$
\begin{equation}
{\bf{M}} = -\frac{\partial F}{\partial {\bf{H}}_0} \ .
\label{magnet}
\end{equation}
The free energy $F$ is related to the partition function $Z$ by 
\[
F=-kT \log Z \quad {\text{with}} \quad Z={\text{Sp}}e^{-\beta H} \quad
{\text{and}} \quad \beta=\frac{1}{k_B T} \; ,
\]
i.e. for the Hamiltonian (\ref{hamdiag})
\begin{equation}
F=E_{cl}+\frac{1}{2} \sum_{\bf{q}} (E_{\bf{q}}-A_{\bf{q}}) + k_B 
T \sum_{\bf{q}} \log (1-e^{-\beta E_{\bf{q}}}) \;.
\label{freeenergy}
\end{equation}
Inserting Eq. (\ref{freeenergy}) into (\ref{magnet}) yields for the
magnetization 
\begin{equation}
{\bf{M}}={\bf{M}}_{cl}-{\bf{M}}_0-{\bf{M}}_{th}(T) \; ,
\label{magnetisation}
\end{equation}
which consists of a classical part 
\begin{equation}
{\bf{M}}_{cl} = -\frac{\partial E_{cl}}{\partial {\bf{H}}_0} \;,
\label{mclass}
\end{equation}
a part arising from quantum fluctuations ($T=0$)
\begin{equation}
{\bf{M}}_0=\frac{1}{2} \sum_{\bf{q}} \left\{
\frac{1}{E_{\bf{q}}} \left(
B_{\bf{q}} \frac{\partial B_{\bf{q}}}{\partial {\bf{H}}_0}
- A_{\bf{q}} \frac{\partial A_{\bf{q}}}{\partial {\bf{H}}_0} \right)
+\frac{\partial A_{\bf{q}}}{\partial {\bf{H}}_0} \right\}
\label{mquant}
\end{equation}
and a part arising from thermal fluctuations
\begin{equation}
{\bf{M}}_{th}(T)=\sum_{\bf{q}} \frac{1}{e^{\beta E_{\bf{q}}}-1}
\frac{1}{E_{\bf{q}}} \left(
B_{\bf{q}} \frac{\partial B_{\bf{q}}}{\partial {\bf{H}}_0}
- A_{\bf{q}} \frac{\partial A_{\bf{q}}}{\partial {\bf{H}}_0} \right) \;.
\label{mthermisch}
\end{equation}
The classical magnetization is reduced by quantum and thermal fluctuations. 
For the spin-flop phase the classical magnetization can be obtained by 
inserting the expression for the classical ground state energy 
(Eq. (\ref{eclassicx})) in Eq. (\ref{mclass}). We get
\begin{equation}
M_{cl}^x = g \mu_B N S \sin \varphi \;,
\label{mclx}
\end{equation}
where $\sin \varphi$ is given by Eq. (\ref{phix}).
The contributions of quantum and thermal fluctuations will be taken into
account numerically when we compare our results with experiments 
(s. section IV).

Now we turn to the staggered magnetization in the $z$-direction 
\begin{equation}
{\text{N}}_z=g \mu_B \sum_l \sigma_l \langle S_l^z \rangle \;,
\label{nz}
\end{equation}
which leads to 
\begin{equation}
{\text{N}}_z=g \mu_B \cos \varphi \left( 
NS - \frac{1}{2} \sum_{\bf{q}} \left( \frac{A_{\bf{q}}}{E_{\bf{q}}}
-1 \right) - \sum_{\bf{q}} \frac{A_{\bf{q}}}{E_{\bf{q}}} n_{\bf{q}} \right) \,.
\label{altmag1}
\end{equation}
The second and third term in (\ref{altmag1}) are divergent due to the
Goldstone-mode. Thus, no antiferromagnetic long-range order is possible in this
case, only a finite magnetization.

\subsection{Field perpendicular to the chain direction}

\subsubsection{Exchange-dominated region $(\kappa < \kappa_c)$}

A field perpendicular to the chain leads to a spin-flop-transition; for
infinitesimal fields the spins lie in the $yz$-plane and perpendicular to the
applied field. With increasing field the spins become progressively aligned in
field direction. In the following calculations the field is applied in the
$y$-direction so that for infinitesimal field the spins are oriented in the
$\pm z$-direction.  Again, we study the classical
\begin{figure}[htb]
 \epsfysize=0.9\columnwidth\rotate[r]{ 
  \epsfbox{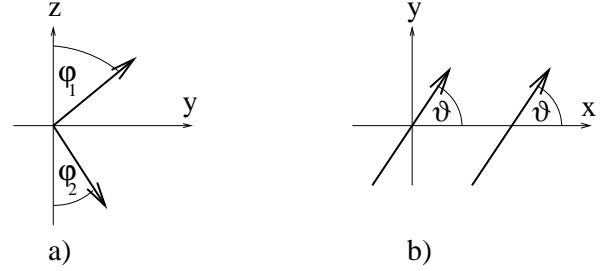}}
\mbox{}  \hspace{4cm}
  \caption{Configuration of the spins for a field applied perpendicular to the
  chain axis; a) for the exchange-dominated region,
  $\kappa < \kappa_c$, b) for $\kappa > \kappa_c$.}
   \label{flopz}
\end{figure}
\noindent
ground state energy of a
general two-sublattice system, for which the usual spin-flop phase turns out to
be stable. The equilibrium condition leads to
\begin{equation}
\sin \varphi = \frac{g \mu_B H_0^y}{2S(J_{{\bf{q}}_0}-J_0
+A_{{\bf{q}}_0}^{33}-A_0^{22})} \; , \label{phiz}
\label{winkelflopz}
\end{equation}
in analogy to Eq. (\ref{phix}) for fields along the chain axis, which
differs only in the dependence on the dipole interaction. 
The transition between the
spin-flop phase and the paramagnetic phase occurs at (see Fig. \ref{phasez})
\begin{equation}
h_{c}^y (\kappa) = 8 + \kappa \bigl( 2 \eta(3) +  2 \zeta(3) \bigr) 
\quad {\text{for}} \quad \kappa<\kappa_c \; .
\end{equation} 
The coefficients of the Hamiltonian, Eq. (\ref{hamvordiag}), in the spin-flop
phase are in linear spin-wave theory
\begin{eqnarray}
A_{\bf{q}}&=& S(2J_{{\bf{q}}_0}-J_{{\bf{q}}+{\bf{q}}_0}-J_{\bf{q}}
+ 2 A_{{\bf{q}}_0}^{33} - A_{{\bf{q}}+{\bf{q}}_0}^{22} 
- A_{\bf{q}}^{11})\nonumber\\
&&+S \sin^2 \varphi \, 
(J_{{\bf{q}}+{\bf{q}}_0}-J_{\bf{q}}+A_{{\bf{q}}+{\bf{q}}_0}^{22}
-A_{\bf{q}}^{33}) \;, \nonumber\\
B_{\bf{q}}&=& S(J_{{\bf{q}}+{\bf{q}}_0} - J_{\bf{q}} + 
A_{{\bf{q}}+{\bf{q}}_0}^{22}
-A_{\bf{q}}^{11})\nonumber\\
&&+S \sin^2 \varphi \, (J_{\bf{q}}-J_{{\bf{q}}+{\bf{q}}_0}
+A_{\bf{q}}^{33}-A_{{\bf{q}}+{\bf{q}}_0}^{22}) \;.
\label{aqbqz}
\end{eqnarray}
For finite $\kappa$ rotational symmetry is lost for the spin configuration
of Fig. \ref{flopz}a and thus no Goldstone-mode is present in the dispersion
relation of Fig. \ref{dispz} showing the spectrum for
two different values of $\kappa$ and $H_0^y$.
\begin{figure}[htb]
  \narrowtext
 \epsfxsize=0.6\columnwidth\rotate[r]{
  \epsfbox{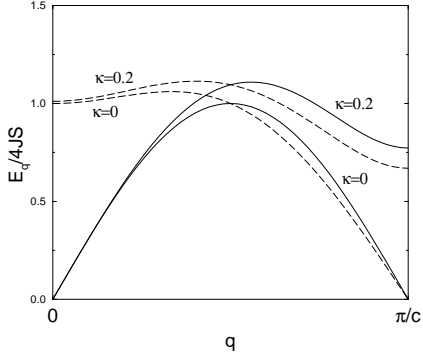}}
\mbox{}
\hspace{5cm}
  \caption{Dispersion relation in the spin-flop phase for
  different magnetic fields applied perpendicular to the chain direction and
  different strengths of the dipolar interaction. The solid lines are without
field and the dashed lines are for $\sin \varphi=0.5$
(Eq. ({\protect{\ref{phix}}})).}
\label{dispz}
\end{figure}
The magnetization in the $y$-direction is obtained from the free energy 
(see Eq. (\ref{magnet})) and is of the same form as
Eqn. (\ref{magnetisation})-(\ref{mclx}) with different canting
angle $\sin \varphi$ (Eq. (\ref{phiz})) and different 
coefficients $A_{\bf{q}}$ and $B_{\bf{q}}$ (Eq. (\ref{aqbqz})).
 
Now the staggered magnetization in $z$-direction, which in analogy to 
Eqs. (\ref{nz})-(\ref{altmag1}) is given by
\begin{equation}
{\text{N}}_z=g \mu_B \cos \varphi \left( 
NS - \frac{1}{2} \sum_{\bf{q}} \left( \frac{A_{\bf{q}}}{E_{\bf{q}}}
-1 \right) - \sum_{\bf{q}} \frac{A_{\bf{q}}}{E_{\bf{q}}} n_{\bf{q}} \right) 
\, ,
\end{equation}
is finite, because of the energy gap in $E_{\bf{q}}$
(s. Fig. \ref{dispz}). However, we believe that due
to nonlinear kink-like excitations again no long-range order is possible for
finite temperatures.

If the field is applied in chain direction a Goldstone-mode remains
(s. Fig. \ref{dispx}) and hence the expression for the staggered
magnetization ${\text{N}}_z$ diverges (Eq. (\ref{altmag1})).

\subsubsection{Dipole dominated region ($\kappa > \kappa_c$)}

\begin{figure}[htb]
  \narrowtext   
 \epsfxsize=0.6\columnwidth\rotate[r]{
  \epsfbox{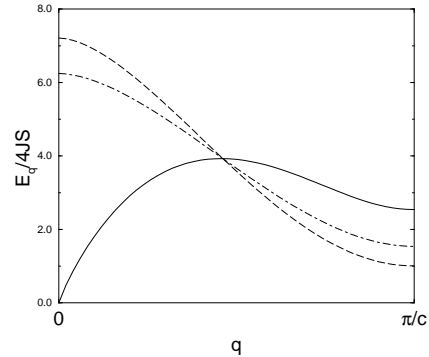}}
\mbox{}
\hspace{5cm}
\caption{Dispersion relation for the dipole dominated region with a field
  perpendicular to the chain direction and $\kappa=4$; dashed: $\sin
\vartheta=0$, dot-dashed: $\sin \vartheta=0.5$ and solid: $\sin \vartheta=1$.}
\label{ferrokant.fig}
\end{figure}

\begin{figure}[htb]
  \narrowtext
 \epsfxsize=0.6\columnwidth\rotate[r]{
  \epsfbox{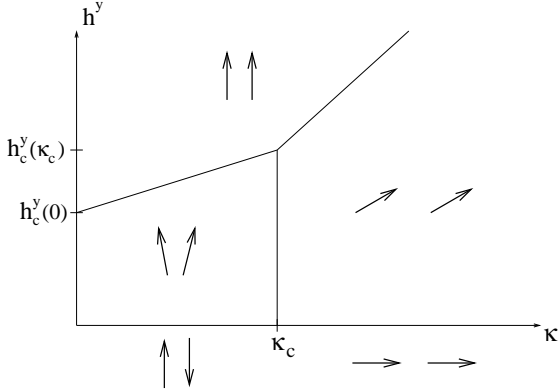}}
\mbox{}
\hspace{4cm}
  \caption{Magnetic phase diagram of the dipolar, antiferromagnetic chain with
  a field perpendicular to the chain axis. The configurations below
  the $x$-axis correspond to vanishing field.}
   \label{phasez}
\end{figure}

In this region the spins keep the ferromagnetic orientation but with a canting
angle $\vartheta$ (intermediate phase) with respect to the magnetic field (s.
Fig. 7b). The classical ground state energy is given by
\begin{eqnarray}
E_{cl}&=&- NS^2 \left[ (J_0+A_0^{22}) \sin^2 \vartheta
+ (J_0+A_0^{11}) \cos^2 \vartheta \right] \nonumber\\
&& -g \mu_B H_0^y N S \sin \vartheta \;,
\label{eintermediate}
\end{eqnarray}
the minimization of which gives the following relation between the canting 
angle $\vartheta$ and the applied field $H_0^y$: 
\begin{equation}
\sin \vartheta = \frac{g \mu_B H_0^y}{2S(A_0^{11}-A_0^{22})} \;.
\label{winkelintermediate}
\end{equation}
Note that this angle depends only on the dipole strength but not on the
exchange energy due to the isotropy of $J_{\bf q}$. The transition 
to the paramagnetic phase, in which all spins orient along the magnetic field,
occurs at
\begin{equation}
h_{c}^y (\kappa) = 6 \zeta(3) \kappa \quad {\text{for}} \quad
\kappa>\kappa_c \; .
\end{equation}
The coefficients in the Fourier transformed 
Hamiltonian for finite canting angle $\vartheta$ are given by 
\begin{eqnarray}
A_{\bf{q}}&=& 2S(J_{0}-J_{\bf{q}})
+ 2S(A_0^{11} - A_{\bf{q}}^{22}) \nonumber \\ 
&&+ S \sin^2 \vartheta \, (A_{\bf{q}}^{22}-A_{\bf{q}}^{11})\nonumber\\
B_{\bf{q}}&=& S \sin^2 \vartheta \, (A_{\bf{q}}^{22}-A_{\bf{q}}^{11}) \;.
\label{ferro}
\end{eqnarray}
The spectrum is shown for three values of the field in 
Fig. \ref{ferrokant.fig}. For decreasing field the energy gap 
$E_{{\bf{q}}_0}$ decreases until it vanishes at the transition point.
Now we investigate the transition from the spin-flop phase to the
intermediate phase as a function of $\kappa$. Comparison of the
classical ground state energy of the two phases shows that the transition line
is independent of the applied field $H_0^y$ (if $H_0^y < H_c^y$), which 
explains the vertical line at $\kappa=\kappa_c$ 
between the spin-flop phase and the
intermediate phase in the phase diagram of Fig. \ref{phasez}. 
For $\kappa < \kappa_c$ the spin-flop phase has lower
energy and is thus the stable phase, for $\kappa > \kappa_c$ the intermediate
phase has lower energy. At $\kappa=\kappa_c$ and $h=h_c$ the energies of the
spin-flop phase, the intermediate phase and the paramagnetic phase are the
same.

\section{Comparison with experiment}

\begin{figure}
  \narrowtext
 \epsfxsize=0.8\columnwidth\rotate[r]{
  \epsfbox{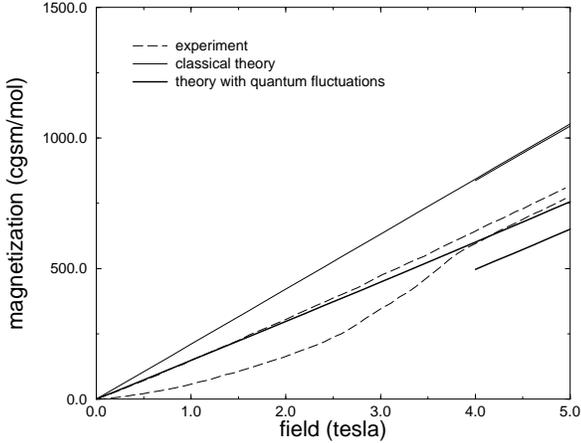}}
\mbox{}
\hspace{4cm}
\caption{Magnetization curves for RbMnBr$_3$
{\protect{\cite{abanov}}}: The dashed lines show the
  experimental measurements at $T=1.7$K.
  The thin lines show the classical results of our
  theory, i.e. without quantum fluctuations. The thick lines show the results
  taking into consideration quantum fluctuations. For each pair of curves the
  upper one refers to field parallel and the lower one to field perpendicular
to the chain.}
\label{theoexp}
\end{figure}

In this section we compare our theoretical results with measurements on quasi
one-dimensional antiferromagnets in the hexagonal ABX$_3$ structure. Due to
different exchange mechanisms the intrachain interaction is about three orders
of magnitude larger than the interchain interaction, thus that these materials
behave in a quasi one-dimensional manner. The most promising substances are the
manganese compounds, because Mn$^{2+}$ ions have spin $S=5/2$ and vanishing
orbital momentum. Thus, the main contribution to the anisotropy should be the
dipole-dipole interaction. The field-dependent magnetization has been measured
for RbMnBr$_3$ (Fig. \ref{theoexp} dashed lines) \cite{abanov}.  In this
material the spins are forced to orient within the hexagonal plane,
perpendicular to the chain-axis. This planar anisotropy can be fully explained
by the dipole-dipole interaction, which leads in the exchange-dominated region
to just such an orientation (see section II). Magnetization experiments show an
anisotropy: the magnetization for magnetic fields along the spin-chain axis is
about 5 to 10\% larger than for fields perpendicular to it. Part of this
anisotropy can be understood qualitatively even for classical spins by the
dipolar anisotropy, which leads to different dependences of the angle $\varphi$
of the spins on the magnetic field (Eqs. (\ref{phix}) and (\ref{phiz})).  The
classical contribution to the magnetization is larger for parallel than
perpendicular field, since the dipole interaction favors ferromagnetic
orientation in chain direction.  For a quantitative comparison we need the
values of the exchange energy $J$ and the lattice constant $c$ which are given
for RbMnBr$_3$ by \cite{Heller}
$$
J = 9.56 K, \quad c = 3.26 {\rm \AA} \; .
$$
Here the exchange energy was determined by a fit of the spin-wave curves with
the spin-wave spectrum calculated from a Hamiltonian in which the anisotropy is
described by a single-ion term. For this Hamiltonian there exists a
renormalization calculation \cite{Khvesh}, which shows, that the classical 
value of the exchange energy has to be renormalized due to magnon-magnon 
interactions. The resulting value of $J$ is
$$
J=8.93 K \;.
$$
This corresponds to $\kappa=8.07 \cdot 10^{-3}$, which is deep in the
exchange dominated region of our theory. With this constant we calculate the
classical contributions to the magnetization (see Eq. (\ref{mclx})) in the
spin-flop phase shown in Fig. \ref{theoexp} as the upper and lower thin solid
lines for fields along and perpendicular to the chain-axis. The dashed lines
represent the experimental values. The transition to the paramagnetic phase
takes place at $H_c \approx 130 T$ and cannot be seen here. For perpendicular
fields lower than $H \approx 3.9 T$ the three-dimensional ordering becomes
important: the system changes to a complicated six-sublattice structure
\cite{Chubukov} for which our one-dimensional model is inappropriate.
According to Fig. \ref{theoexp}, the classical magnetization is larger than the
experimental and the splitting for the two directions is almost invisible.
However, in one dimension the quantum fluctuations are strong and must be taken
into account. Numerical evaluation of the magnetization for longitudinal and
perpendicular field direction (Eq. (\ref{magnetisation})), including the
zero-point contributions ($T=0$), yields the thick solid lines in Fig.
\ref{theoexp}. In fact, the quantum fluctuations lead to a pronounced reduction
of the magnetization and increase the splitting, but as to be expected, are
somewhat overestimated in the pure one-dimensional model.  The contribution of
thermal fluctuations to the magnetization (Eq. (\ref{mthermisch})) is
negligible for temperatures as low as in the experiment ($T=1.7 K$). This is in
contrast to the work of Santini et al. \cite{Santini}, who claim that the
classical thermal fluctuations are the relevant contribution and totally
neglect the quantum fluctuations.  Abanov and Petrenko \cite{abanov} also
studied the one-dimensional spin-chain, but with a single-ion anisotropy
favoring orientation perpendicular to the chain instead of the dipolar
interaction. Inserting the fitted value for this anisotropy they get a larger
fluctuation contribution than in our theory. Furthermore, the classical
contribution shows the opposite anisotropy, i.e. the magnetization for a field
along the chain-axis is lower than for a perpendicular field.

\begin{figure}[htb]
 \epsfxsize=0.6\columnwidth\rotate[r]{
  \epsfbox{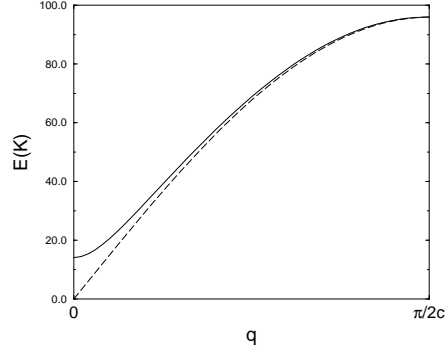}}
 \hspace{5cm}
\caption{The two branches of the dispersion relation for RbMnBr$_3$ in the   
magnetic Brillouin zone with dipolar interaction ($\kappa=8.072 \cdot 10^{-3}$,
$J=9.56$ K).}
\label{disprbmn}
\end{figure}

In summary, we can explain the anisotropy measured in RbMnBr$_3$
qualitatively by a one-dimensional, dipolar Heisenberg model. We expect that
by taking into account the full three-dimensional structure with interactions
between different chains the quantitative agreement could be improved because
the contribution of quantum fluctuations is reduced in higher dimensions.
Our prediction for the spin wave dispersion relation of RbMnBr$_3$ (Eq.
(\ref{magspectrum})) is shown in the magnetic Brillouin zone in Fig.
\ref{disprbmn}. Here we set the exchange interaction $J$ to $9.56 \, K$. The
resulting gap of the upper dispersion branch at ${\bf{q}}=0$ equals 
$14.16 \, K$.
\section{Summary}
We have considered the one-dimensional, isotropic Heisenberg antiferromagnet
with dipole-dipole interaction in the complete parameter region. We showed that
there exists an exchange dominated phase for $\kappa < \kappa_c$, where the
spins orient antiferromagnetically, perpendicular to the chain axis, i.e. the
dipole-dipole interaction gives rise to a planar anisotropy. For $\kappa >
\kappa_c$, the dipole-dominated phase, the ground state is ferromagnetic along
the chain axis, i.e. the dipole-dipole interaction acts like an easy-axis
anisotropy. Because of the broken rotation symmetry the dipole-dominated region
shows long-range order for $T=0$. In the exchange-dominated region there is
still a Goldstone-mode due to the rotation symmetry around the chain axis.
Although the linear small ${\bf{q}}$ behavior of the acoustic magnon mode is
modified by the dipole energy (Eq. (\ref{entw})), there is no long-range order
neither for finite nor for zero temperature. 

The magnetic phase diagram for fields along and perpendicular to the chain in
both the dipole and the exchange dominated regions has been determined and the
predicted magnetization has been compared with experiments on RbMnBr$_3$, a
quasi one-dimensional, planar antiferromagnet. The experimental anisotropy can
be explained qualitatively by the one-dimensional, dipolar model, the
quantitative agreement could be improved by a more elaborate, three-dimensional
calculation.

\acknowledgments
This work has been supported by the BMBF under contract number 03-SC4TUM.
One of us (M.H.) has benefited from a scholarship of the Studienstiftung des
Deutschen Volkes. The work of C.P. has been supported by the Deutsche
Forschungsgemeinschaft (DFG) under the contract no. PI 337/1-1.

\appendix
\section{Dipolar sums}
\begin{figure}  
  \narrowtext
 \epsfxsize=0.6\columnwidth\rotate[r]{
  \epsfbox{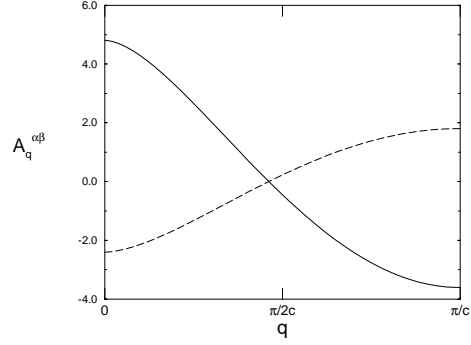}}
\mbox{}
\hspace{4cm}
\caption{Dipole sums $A_{\bf{q}}^{\alpha \beta}$ in Fourier space given in
units of $\frac{{(g \mu_B)}^2}{2c^3}$. The solid line denotes
$A^{11}_{\bf{q}}$ and the dashed $A^{22}_{\bf{q}}=A^{33}_{\bf{q}}$.
All non-diagonal components vanish.}
\label{fig12}   
\end{figure}

The dipolar sums can be evaluated in Fourier space by the Ewald 
method \cite{Bonsall}. The sum in Fourier space is divided into a sum over 
the direct lattice and a sum over the indirect lattice, so that both sums
converge quickly.
The dipolar tensor in one dimension finally reads (all values of the dipolar 
tensor are measured in units of $\frac{{(g \mu_B)}^2}{2 c^3}$)
\begin{equation}
A^{\alpha \beta}_{\bf{q}}
= \left\{ \begin{array}{lr}
\frac{4}{3} \pi
- 2 \pi \sum_{l}^{'} e^{i{\bf{q}}{\bf{x}}_{l}} 
\varphi_{\frac{1}{2}}(\frac{\pi}{c^2} {| {\bf{x}}_{l} |}^2)\\ 
-2 \pi \sum_{r} \varphi_{-2} 
\left( \frac{c^2}{4 \pi} {| {\bf{q}}+{\bf{G}}_r |}^{2} \right)
\quad \quad \alpha = \beta \; \not \in \; {\bf{x}}_l\\
 \\
\frac{4}{3} \pi \delta_{\alpha \beta} - 2 \pi \delta_{\alpha \beta}
\sum_{l}^{'} e^{i{\bf{q}}{\bf{x}}_{l}} \varphi_{\frac{1}{2}}
(\frac{\pi}{c^2}{| {\bf{x}}_{l} |}^2) \\
+ 4 \frac{{\pi}^2}{c^2} 
\sum_{l}^{'} e^{i{\bf{q}}{\bf{x}}_{l}} x_l^{\alpha} x_l^{\beta}
\varphi_{\frac{3}{2}}(\frac{\pi}{c^2} {| {\bf{x}}_{l} |}^2)\\
-c^2\sum_{r} {({\bf{q}}+{\bf{G}}_r)}_{\alpha} {({\bf{q}}+{\bf{G}}_r)}_{\beta}\\
\qquad \quad \times \;
\varphi_{-1} \left( \frac{c^2}{4\pi}{| {\bf{q}}+{\bf{G}}_r |}^{2} \right)
\hfill \alpha, \beta \in {\bf{x}}_l \;,
\end{array}
\right.
\label{dipsum1d}
\end{equation}
where ${\bf{x}}_l=l c(1,0,0)$ and ${\bf{G}}_r=\frac{2\pi}{c} r (1,0,0)$.
The prime at the $l$-summations denotes $l \neq 0$.
We have to distinguish between components with spins lying in the 
chain direction ($\alpha,\beta \in {\bf{x}}_l$) and those perpendicular to it
($\alpha, \beta \: \not \in  \, \: {\bf{x}}_l$). The off-diagonal components
are identical zero due to the inversion symmetry. The dipolar sums are shown in
Fig. (\ref{fig12}) as a function of ${\bf{q}}$. 
In Eq. (\ref{dipsum1d}) we have introduced the 
Misra functions
\begin{equation}
\varphi_{n}(x)=\int_{1}^{\infty} dt \, t^n \, e^{-xt} \;.
\end{equation}
The asymptotic behavior of the Misra functions for small arguments $x$ is
\begin{equation}
\varphi_{-1}(x) \approx  (-\gamma - \log x) + x -\frac{1}{4} x^2 +
{\cal{O}}(x^3) 
\end{equation}
and
\begin{equation} 
\varphi_{-2}(x) \approx 1 + (-1 + \gamma + \log x) x -\frac{1}{2} x^2 +
{\cal{O}}(x^3) \;,
\end{equation}
where $\gamma \approx 0.5772$ is Euler's constant. The expansion of the dipolar
tensor needed in Eq. (\ref{entw}) reads 
\begin{eqnarray}
A_{\bf{q}}^{11}& =& c_1 - c_2 {(qc)}^2 + {(qc)}^2 \log \left( 
\frac{{(qc)}^2}{4\pi} \right) \nonumber\\
&&+c_3 {(qc)}^4 + {\cal{O}} \left( {(qc)}^6 \right) 
\end{eqnarray}
\begin{equation}
A_{{\bf{q}}+{\bf{q}}_0}^{22} = A_{{\bf{q}}+{\bf{q}}_0}^{33}
= c_4 - c_5 {(qc)}^2 + c_6 {(qc)}^4 +{\cal{O}} \left( {(qc)}^6 \right)
\; ,
\end{equation}
where the constants $c_1, c_4$ are given analytically in
Eq. (\ref{a11exakt}) and (\ref{a33exakt}), while the others are
computed numerically as
\begin{equation}
\label{cvalues}
c_2 \approx 0.666 \, , \; c_3 \approx 0.066 \, , \; c_5 \approx 0.693
\quad {\text{and}} \quad c_6 \approx 0.021 \; .
\end{equation}
The dipolar tensor can be evaluated
exactly for some configurations. For the ferromagnetic order of the spins 
along the $x$-axis one gets
\begin{equation}
A^{11}_0 = 4 \sum_{l>0} \frac{1}{l^3}= 4
\zeta (3) = c_1 \approx 4.808 \;,
\label{a11exakt}
\end{equation}
where $\zeta(n)$ is the Riemann zeta function\cite{Abram}
\begin{equation}
\zeta (n) = \sum_{l=1}^{\infty} l^{-n} \;.
\end{equation}
The antiferromagnetic order of the spins along the $z$-axis yields
\begin{equation}
A^{33}_{{\bf{q}}_0}= 2 \sum_{l>0} 
\frac{{(-1)}^{l-1}}{l^3}=2 \eta(3) = c_4 \approx 1.803
\label{a33exakt}
\end{equation}
where the eta function $\eta(n)$ is defined by\cite{Abram}
\begin{equation}
\eta (n) = \sum_{l=1}^{\infty} {(-1)}^{l-1} l^{-n} \;.
\end{equation}
Other exact values are
\begin{equation}
A^{33}_0=-2 \zeta(3) \quad {\text{and}} \quad 
A^{11}_{{\bf{q}}_0}=-4 \eta(3) \;.
\end{equation}
These values corroborate the sums in Eq. (\ref{dipsum1d}).
One can show that for the dipolar tensor components of the linear chain the
relation
\begin{equation}
A^{11}_{{\bf{q}}}+A^{22}_{{\bf{q}}}+A^{33}_{{\bf{q}}}=
A^{11}_{{\bf{q}}}+2 A^{22}_{{\bf{q}}}=0
\end{equation}
is valid.

The constants in Eq.(\ref{entw}) are given by
\begin{equation}
b_1=\frac{1}{16} \Bigl( 8-\kappa \bigl( 4 \zeta(3) -2 \eta (3) \bigr) \Bigr)
\Bigl( 2+\kappa c_5 \Bigr)
\label{bvalue1}
\end{equation}
\begin{eqnarray}
b_2=-\frac{1}{3} + \kappa &\Bigl(
&\frac{1}{96}\bigl(4 \zeta(3) -2 \eta (3) \bigr) +\frac{1}{8}
(c_2-c_5)-\frac{1}{2}c_6\nonumber\\
& + &\frac{1}{16} \kappa c_2 c_5 + \frac{1}{16} 
\kappa c_6 \bigl(4 \zeta(3) -2 \eta (3)\bigr) \nonumber\\
& + &\frac{1}{8} \log (4 \pi)
+\frac{1}{16} \kappa c_5 \log (4 \pi )\Bigr) \; .
\label{bvalue2}
\end{eqnarray}

%\pagebreak


\begin{references}
\bibitem{pich2000} It can be shown that the ferromagnetic orientation is 
the exact ground state of the dipolar chain, C. Pich, to be published.
\end{references}
\end{document}